\documentclass[preprint]{epl}

\title{Segregation by friction}
\author{L. Kondic\inst{1} 
	\and R.~R. Hartley\inst{2} 
	\and S.~G.~K. Tennakoon\inst{2}
	\and B. Painter\inst{2}
	\and R.~P. Behringer\inst{2}}
\shortauthor{L. Kondic \etal}
\institute{
  \inst{1} Department of Mathematical Sciences \& Center for Applied
  Mathematics \& Statistics\\ New Jersey Institute of Technology,
  Newark, NJ 07102\\
  \inst{2} Department of Physics and Center for Nonlinear and Complex
  Systems\\ Duke University, Durham NC, 27708-0305 }


\begin{document}

\maketitle

\begin{abstract}
Granular materials are known to separate by size under a variety of
circumstances.  Experiments presented here and elucidated by modeling
and MD simulation document a new segregation mechanism, namely
segregation by friction.  The experiments are carried out by placing
steel spheres on a horizontal plane enclosed by rectangular sidewalls,
and subjecting them to horizontal shaking.  Half the spheres are
highly smooth; the remainder are identical to the first half, except
that their surfaces have been roughened by chemical etching, giving
them higher coefficients of friction.  Segregation due to this
difference in friction occurs, particularly when the grains have a
relatively long mean free path.  In the presence of an appropriately
chosen small ``hill'' in the middle of the container, the grains can
be made to completely segregate by friction type.
\end{abstract}

Granular materials have been the subject of much recent as well as
past attention~\cite{reviews1,reviews2,reviews3}.  A steady input of
energy in the presence of gravity, for instance by shaking, causes
materials of different sizes to segregate, with the largest grains
tending toward the top.  This phenomenon is called the ``Brazil Nut
Effect''.
Explanations~\cite{exp1,exp2,exp3,exp4,hill_95,exp6,alexander_01,jenkins_02}
involve various geometric arguments as well as flows due to
wall-driven convection.  While other sources of segregation and/or
clustering have been studied~\cite{exp5,eggers_99}, a much less
explored issue is the role played by friction in segregation.  Here,
we investigate the issue of friction-related segregation in a regime
where a two-dimensional granular material is relatively
fluid-like~\cite{gas1,gas2,gas3}, and in an apparatus where wall
effects are generally not important.  Although the material may be
roughly analogous to a fluid in these experiments, we observe a kind
of segregation that does not occur in a true thermodynamic system.

\begin{figure}
\onefigure[width=12cm]{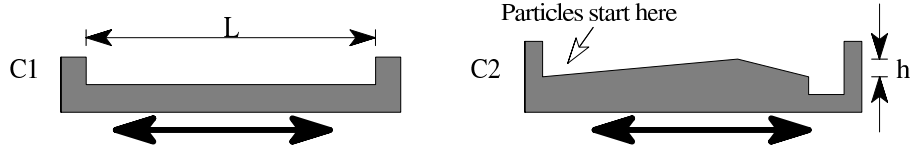}
\caption{Sketch of the flat shaken surface (C1) and the hill
configuration (C2).  Here, $L= 28.73$ cm, $h = 0.22$ cm, and the
dimension of the system in the transverse direction (not shown) is
$11.23$ cm.  }
\label{fig:exp_geom}
\end{figure}

The geometry for these experiments is sketched in
fig.~\ref{fig:exp_geom}.  We use an electromagnetically driven
shaker~\cite{tennakoon_98} to provide a sinusoidal horizontal
displacement, $x = A \sin (\omega t)$.  Here, A and $\omega$ provide
useful measures for length and time, and the dimensionless
acceleration is $\Gamma \equiv A \omega^2/g$, where $g$ is the
acceleration of gravity.  Two configurations have been explored.  In
the first, C1, the surface on which the spheres reside is flat,
horizontal, and oscillates horizontally.  In the second, C2, the
surface is bent so that there is a small hill, and all the spheres are
initially at one side of the plane.  If the particles cross over the
hill, they are not allowed to return, and we indicate this
schematically in fig.~\ref{fig:exp_geom} by showing a well on the side
of the hill opposite the starting point.  The main difference between
this setup and the one by Kudrolli et al.~\cite{kudrolli_97} is that
here the energy is supplied to the granular particles by the substrate
and boundaries system, while in~\cite{kudrolli_97} the energy is
supplied only by a moving boundary.

The experiments use a 50-50 mixture of ``rough'' and ``smooth'' steel
spheres of $d \approx 4.05$ mm.  The spheres are identical except for
their static, kinematic and rolling friction coefficients, $\mu_s$,
$\mu_k$, and $\mu_r$, respectively.  The rough spheres ($d$) have been
modified by chemical etching and are darker than the lighter colored
unetched spheres ($l$), and we use this as a simple way to distinguish
the two types in the experiments.  For the smooth particles, $\mu_s =
0.33 \pm 0.04$, $\mu_k = 0.10 \pm 0.03$, and $\mu_r = (1.9 \pm 0.3)
\times 10^{-3}$, while for rough particles, $\mu_s = 0.41 \pm 0.04$,
$\mu_k = 0.12 \pm 0.02$, and $\mu_r = (4.2 \pm 0.6) \times 10^{-3}$.
The frictional properties were measured using the fast camera
techniques given in~\cite{painter}.  Specifically, to determine
$\mu_s$ and $\mu_k$, we glued three particles together and placed them
on an inclined plane.  The tangent of the angle at which the particles
just began to slide defined $\mu_s$; the trajectory of the sliding
particles then yielded $\mu_k$.  Finally, to determine $\mu_r$, we
placed a single particle on a flat substrate, and tracked its motion
from an initially rolling state.  The deceleration then yielded
$\mu_r$.  We emphasize that the measurement techniques have relatively
large error bars.  But, by design, the friction coefficients are
higher for the etched particles.  It is also interesting to note that
segregation occurs for even modest differences in friction, i.e. a
large difference is not needed to obtain the effect.

\begin{figure}
\onefigure[height=8.5cm]{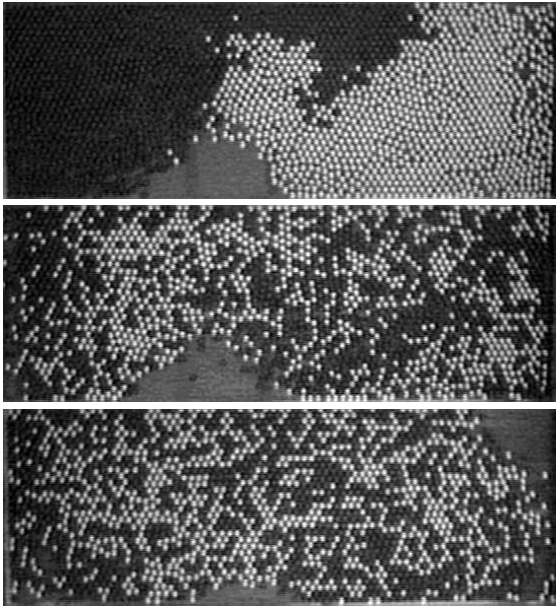}
\caption{Typical pattern of spheres after shaking ($\Gamma = 0.130 \pm
0.005$, $A = 1.80$ cm, and $f=1.339 \mathit{Hz}$) for a horizontal
surface (C1 protocol) after 1, 6 and 16 minutes (top to bottom).  Dark
grains have higher rolling (and other) friction coefficients than
light grains.  The distribution by friction type evolves from almost
completely segregated for $t=1$ min to randomized for $t=16$ min.}
\label{fig:pattern}
\end{figure}

Figure~\ref{fig:pattern} shows the results of the experiments
performed under the C1 protocol.  Here we use a mixture of 1000 smooth
and 1000 rough spheres that fill roughly $80\%$ of the available
surface area of the shaker.  Initially the particles are completely
segregated, with all the $l$ particles to the right, and all the $d$
particles to the left.  Figure~\ref{fig:pattern} shows the
distribution of the particles after $t=1,6,$ and $16$ minutes.  If the
spheres were identical except for color, then we would anticipate that
they would be randomly distributed in space.  A visual inspection may
suggest that for late times the $l$ spheres form chain-like
structures, and that $d$ spheres form separated structures that are
interleaved with the light particles.  To check this visual
impression, we explore quantitatively the color distribution of the
particles.  If the particles are placed randomly without regard to
color, then the probability of finding $n$ nearest neighbor grains of
any given friction type (i.e. color) in a cluster of $N$ (typically
six) particles around a particle of a given type is a simple binomial
distribution: $P(n) = (1/2)^N N!/(n!(N-n)!)$.  Careful examination of
this probability, using the data given in fig.~\ref{fig:pattern},
shows that the visual impression of structure is an illusion: the
spheres are completely mixed for late times, and they closely follow a
binomial distribution.  Alternatively, if an experimental image of the
spheres is processed so that the colors are reassigned randomly, it is
not possible to distinguish an arbitrary original image from an
arbitrary processed image.  Thus, in the dense region of the sample,
in the absence of a gravitational bias, mixing remains complete.  We
note, however, that a large number of shakes (hundreds) is needed to
completely mix the system (compare fig.~\ref{fig:pattern} for $t=6$
and $t=16$).  Similar results have been observed using different
$\Gamma$'s.

We continue exploring the possibility of segregation by friction using
the C2 protocol.  In this experiment, the spheres (a $50/50$ mixture
of smooth and rough) are all initially placed in a uniformly mixed
state on the left side of the hill.  We can envision spheres crossing
the crest of the hill as being analogous to a thermally activated
process.  If we ignore friction, the probability of a particle 
getting over the hill has the form $P = \exp (-E_b/E_T)$.  
Here $E_b = mgh$ is a measure of the
gravitational energy barrier.  $E_T$ represents a random or
``thermal'' energy, the so-called granular temperature.  Assuming that
the grains are rolling (we will see later that this is not always the 
case), they experience an energy loss per distance
traveled of $dE_T/dx = -\mu_r mg$ that is higher for rough spheres.
Then, if the particles start at the bottom with a nearly identical
random energy, $E_{To}$, the spheres reaching the barrier have a
reduced $E_T$ of $E_T = E_{To} - \mu_r mgx$, where $x$ is the typical
distance traveled to reach the crest of the hill.  In the case of the
rougher spheres, it appears to be possible to adjust the energy input
so that $E_T = 0$ by or before the point that a typical sphere has
reached the crest of the hill.  In that event, the cross-over
probability falls to zero, and segregation is complete.  In practice,
we choose $\Gamma$ so that all $l$ particles cross to the other side
of the hill, whereas very few of $d$ particles cross.  

\begin{figure}
\onefigure[width=7.5cm]{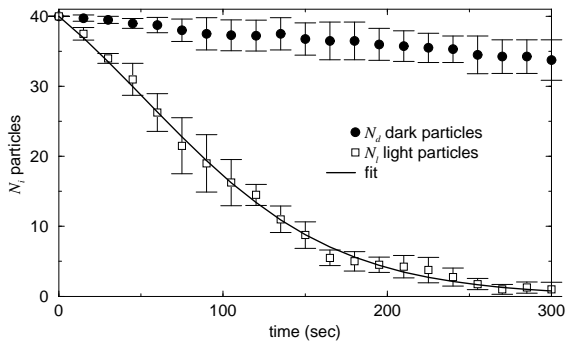}
\caption{The populations of higher- and lower-$\mu$
spheres, $N_d$ and $N_l$, vs. time for the ``hill'' (C2)
configuration.  The line is the least squares fit based on the simple
model discussed in the text, giving $b=0.01737$ and $c=0.0003117$.  }
\label{fig:hill}
\end{figure}

Figure~\ref{fig:hill} shows the results for the number of each type of
sphere, $N_i$ with $i = l,d$, remaining on the starting side of the
hill as a function of time.  This figure shows the mean and standard
deviation of four different experiments.  For simplicity, the $\Gamma$
used here is also the one in fig.~\ref{fig:pattern}, namely, $\Gamma =
0.130$.  The final approach to the steady state (i.e. no $l$ particles
on the original side) appears to be exponential, implying that an
escape rate for the $l$ spheres is given by $ -d N_l/dt \propto N_l$.
However, from fig.~\ref{fig:hill}, it is apparent that there is some
initial curvature to the $N_l$ vs. $t$ data, i.e., departure from
exponential decay, which requires further explanation.

This slow-down of the segregation for early times can be understood by
noting that there are two processes involved.  One (slow) part
consists of collision-dominated diffusion of the $l$ particles to the
free surface of the layer close to the bottom of the hill.  This layer
acts like a liquid phase, preventing the $l$ particles from reaching
the less dense phase further up the slope.  The second (fast) part
consists of their `evaporation' i.e.  ejection away from the bulk.
Once free of the bulk, the dominant energy dissipation 
comes from friction with the substrate.  In the dense
collision-dominated regime, the particles are essentially
indistinguishable; but once they are evaporated into the gas phase,
the particles experience different energy loss rates because of the
difference in the friction.  A simple model that adequately describes
the data and captures at least some of the relevant physics assumes
that
\begin{equation}
dN_l/dt = -a(N_l) N_l,
\label{eq:model}
\end{equation}
where $a(N_l)$ is now a coefficient that depends on $N_l$ as $a = b -
cN_l$, where $b$ and $c$ are positive constants.  Roughly, the idea is
that when the number of low friction particles is low, the escape rate
is at its greatest, namely, $b$.  However, when $N_l$ is relatively
high, there is a collective process that reduces the rate of escape.
That is, it is harder for particles captured inside a dense cluster to
escape.  The smaller the cluster, the easier it is for a particle to
escape the cluster and then possibly make it over the hill.  The
solution of this equation
\begin{equation}
N_l (t) = {b \over c + {b - c N_0 \over N_0} e^{b t}}\, ,
\end{equation}
is shown in fig.~\ref{fig:hill} with $b, c$ fitted using the method of
least squares.

\begin{figure}
\onefigure[width=7.5cm]{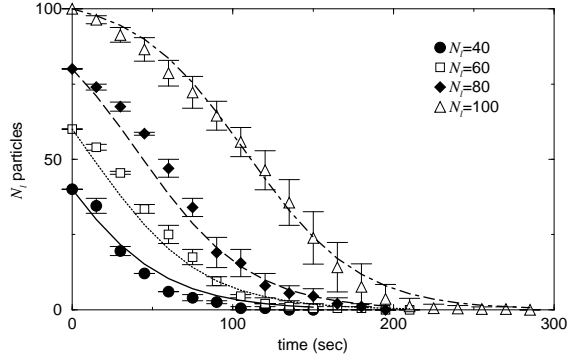}
\caption{The populations of lower-$\mu$ spheres in the C2 experiments
in which higher-$\mu$ spheres are absent.  The lines are the solutions
of Eq.~(\ref{eq:model}) using $b=0.0288$ and $c=0.0002758$.  The
values of the constants $b$ and $c$ are different from those in
fig.~\ref{fig:hill} since there are only $l$ particles in this
system.}
\label{fig:light}
\end{figure}

As an additional test of eq.~(\ref{eq:model}), we have 
performed experiments where only light spheres are present, but their
initial number is varied.  In fig.~\ref{fig:light} we immediately
observe an increased curvature of the profiles for larger
$N$.  This is as expected, since more particles lead to an increased
dissipation of energy, therefore slowing down the escape process.
Based on these results, we conjecture that, while the detailed values
of the parameters depend (weakly) on a particular experimental configuration,
the model specified by eq.~(\ref{eq:model}) captures the most 
relevant physics that governs the escaping probability. 

So far, we have concentrated only on the rolling friction as the
parameter that governs the segregation process.  However, a
back-of-the-envelope energy estimate shows that there is a problem
with this simple picture.  For example, consider a single (rough)
particle moving under gravity, and experiencing collisions with the
left wall shown in the C2 protocol in fig.~\ref{fig:exp_geom}.  It is
easy to see that this particle would be provided with enough energy to
escape to the other side of the hill.  In the case of many particles
there is an additional dissipation due to interparticle collisions,
but still one would expect that during a few hundred cycles, all the
particles (rough and smooth) would escape.  To gain better insight
into this issue and into the segregation mechanism in general, we have
performed discrete element simulations (DES) of this system.

These simulations follow rather closely the model described
in~\cite{kondic99}.  For completeness, a brief overview will be given
here; for details, and for the list of relevant earlier works, the
reader is referred to that paper.  During a collision, a sphere
experiences normal and tangential collisional forces.  The normal
force is given by 
\begin{equation}
{\bf F}_N^c = \left[ k (d - r_{i,j}) - \gamma_N
\bar m ({\bf v}_{i,j} \cdot {\bf\hat n})\right]\, ,
\end{equation}
where $k$ is a force
constant, $r_{i,j} = |{\bf r}_{i,j}|$, ${\bf r}_{i,j} = {\bf r}_i -
{\bf r}_j$, ${\bf \hat n} = {{\bf r}_{i,j}/ r_{i,j}}$, ${\bf v}_{i,j}
= {\bf v}_i - {\bf v}_j$, $\bar m$ is the reduced mass, and $d$ is the
particle diameter.  The damping constant is chosen so that $10$\% of
the energy is lost in a typical collision, as appropriate for steel
particles; see~\cite{kondic99} for other parameters.  We note that
Hertzian (nonlinear) interactions would be appropriate to model the
collision between spheres; however, we have not seen any relevant
difference in the results if the linear model specified above is used,
and the parameters are chosen appropriately~\cite{kondic99}.  The
tangential force in the plane of the substrate is given by \
\begin{equation}
{\bf F}_S^c = sign(-v_{rel}^t) min \left ( \gamma_S \bar m |v_{rel}^t|,
\nu_k |{\bf F}_N^c |\right ) {\bf\hat s}\, ,
\end{equation}
where $v_{rel}^{t}$ is the relative velocity in the tangential
direction ${\bf\hat s}$, $\gamma_S = {\gamma_N/2}$ and $\nu_k$ is the
coefficient of friction between the particles.  In principle, there is
also a tangential collisional force in the direction normal to the
surface, but this is not included here.  Next, the particles interact
with the substrate via rolling friction, as explained earlier.  More
importantly, as shown theoretically~\cite{kondic99}, and measured
experimentally~\cite{painter,painter2}, the particles {\it slide} on
the substrate after a typical collision, due to large momentum
exchange that occurs between the colliding particles.  The basic
picture is that during the collision process, the rotational motion of
the particles cannot immediately follow their translational motion so
as to maintain a no-slip condition.  This leads to a large relative
velocity at the contact point between a colliding particle and the
substrate, and hence to sliding.  Sliding with a coefficient of
friction, $\mu_k$, continues until friction brings the linear and
angular velocities back to a state of rolling without slipping.  The
resulting loss of energy is large, since $\mu_k$ is so much larger
than $\mu_r$.  While the exact loss of energy due to sliding depends
on the details of the collision, and on the time period until the next
collision, this loss could reach $80$ \% of the original energy of a
particle~\cite{painter,painter2,kondic99}.  Therefore, much more
energy can be lost due to sliding, than due to inelasticity of the
collisions.  Sliding also occurs in collisions with the bottom wall.
In principle, sliding could also occur due to acceleration of the
substrate.  However, this is expected only if the acceleration of the
substrates, $a_S$, satisfies $a_S \ge \left( 1+{mR^2/I}\right) \mu_s g
$, where $I$ is the moment of inertia of a particle of radius $R$ and
mass $m$~\cite{kondic99}.  This condition is not satisfied for the
weakly driven case considered here.

Some additional details of the DES are relevant here.  The geometry of
the DE domain is similar (but not exactly identical) to the C2 setup.
In particular, the simulations have been performed by using both
periodic boundary conditions, and rigid (and inelastic) side walls in
the direction transverse to the shaking, without any noticeable
differences.  Initially the particles are placed on a lattice, given
random initial velocities, and left to equilibrate without driving and
without friction.  The particles are then randomly assigned either low
or high friction coefficients, and the simulations are started using
this initial configuration.  We note that for convenience we increase
the inclination angle on the right hand side of the domain in the DES,
so that particles that cross the hill do not return.  In the
experiments, a similar outcome was reached by using a trough and/or
highly absorbent right wall.

\begin{figure}
\onefigure[width=7.5cm]{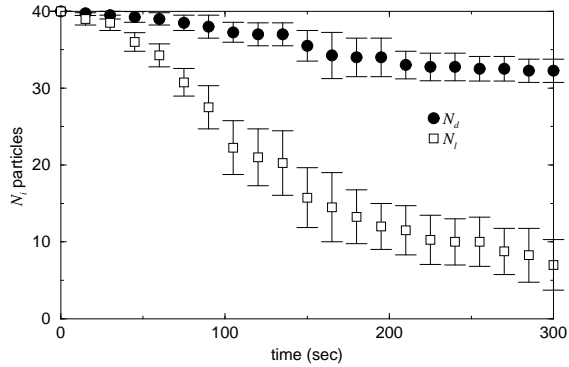}
\caption{The populations of higher- and lower-$\mu$
particles, $N_d$ and $N_l$, vs. time for the ``hill'' configuration
(C2), but in DES.  }
\label{fig:sim}
\end{figure}

Figure~\ref{fig:sim} shows results for the number of remaining particles
vs. time for each type, as obtained from the DES.  These
data were obtained by averaging over four different runs which differ
only in the seed for the random number generator used to assign the
initial particle velocities, as well as their frictional properties.
We show the mean and standard deviation resulting from these four runs.

The escape rates of the DE particles are similar to the ones observed
in experiments, showing that the simulation, while relatively simple,
is complete enough to explain the main features of the experiment.
The only noticeable differences between the simulations and the
experiment are (1) a larger spread of the results of DES compared to
experiments, and (2) somewhat fewer light particles escaping in the
DES.  We note that without inclusion of sliding friction, all the
particles escape to the right side of the domain within the shown time
period.  However, rolling friction is also
important in the segregation process.  For the parameters we explored
in the simulations, we could not achieve a high degree of segregation
based on sliding friction alone (simulations allow for simple removal
of rolling friction by specifying $\mu_r = 0$).  Therefore, both
sliding and rolling frictional properties are relevant to the
segregation process.

It is also worth mentioning that for all the simulations and
experiments carried out here, in the presence of a barrier, less
frictional particles escape across a potential barrier more easily
than those of higher friction.  This observation appears to cover a
broad driving regime, i.e. the particular values given to the
amplitude or frequency of shaking, or of the inclination angle.  This
is in contrast to segregation by size/density, where the outcome may
depend on a shaking regime (see, i.e.~\cite{exp6,jenkins_02}).

To conclude, we have shown that granular systems undergo segregation
due to frictional differences between grain types.   This segregation
process is similar to the case of size differences, although there are
also important differences.  Related frictional segregation may well
occur in other situations when grains are free to flow along a
surface, as in avalanching flows, or possibly in shear
layers near boundaries.  

\acknowledgments This work was supported by NASA grants No. NAG3-2367
and NAG3-2372, and by NSF grants DMR-0137119, DMR-9802602 and
DMS-9803305.  L.K. acknowledges support by NJIT grant No. 421210.

\end{document}